\documentclass[11pt]{article}
\usepackage{moriond,epsfig}

\def\Journal#1#2#3#4{{#1} {\bf #2}, #3 (#4)}

\def\PRD{{\em Phys. Rev.} D}

\def\be{\begin{equation}}
\def\ee{\end{equation}}
\def\bea{\begin{eqnarray}}
\def\eea{\end{eqnarray}}

\begin{document}
\vspace*{4cm}
\title{Single Top production at LHC}

\author{ M.~Mohammadi Najafabadi \\
on behalf of CMS and ATLAS collaborations}

\address{Institute for studies in theoretical physics and mathematics (IPM),\\ 
 School of Physics, P.O. Box 19395-5531, Tehran, Iran}
\maketitle\abstracts{
 The production of single top quarks at LHC provides an ideal framework
  to investigate the properties of the electroweak interaction, in particular of the {\it tWb} coupling.
   Moreover, single top is a powerful mean to identify physics beyond
  the standard model. All three different production mechanisms of single top 
  are expected to be observed at LHC. Recent studies from ATLAS and CMS are presented.}
\section{Introduction}
\begin{figure}[h]
\begin{center}
\epsfig{file=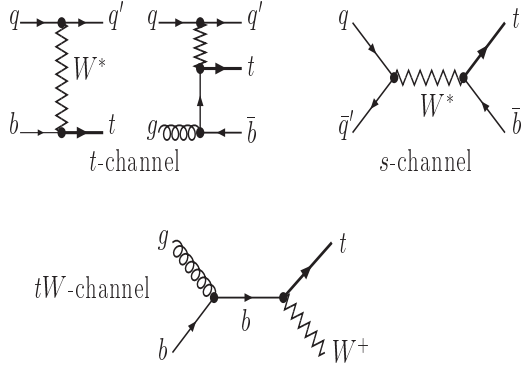,height=5cm,width=7cm}
 \caption{Feynman diagrams for single top production.}
    \label{fig:FeynmanDiagrams}
\end{center}
\end{figure}

Top quarks are mostly produced in pairs via strong interaction at LHC. However, there are
a significant number of top quarks which are produced singly, via the weak interaction.
Single top production proceeds through three separate sub-processes at LHC (Fig.~\ref{fig:FeynmanDiagrams}): 
\begin{itemize}
\item t-channel: the dominant process involves the exchange of a space-like W boson
. This process is also called W-gluon fusion, because the $b$-quark ultimately arises from a 
gluon splitting to $b\bar{b}$.
\item s-channel: involves the production of a time-like W boson, which then decays to a top and a 
bottom quark.
\item tW-channel: involves the production of a real W boson.
\end{itemize}
There are several reasons for studying single top quarks at LHC~\cite{Beneke,Tait}:
\begin{itemize}
\item The cross sections of single top quark processes are proportional to $|V_{tb}|^{2}$. They provide
the only way to measure $V_{tb}$ directly.
\item Single top quarks are produced highly polarized due to V-A structure of weak interaction, so single top provides 
a good laboratory for studying the spin of the top quark.
\item Single top events are backgrounds to new physics signals.
\item New physics could be appeared in single top events. It provides the possibility to investigate the
 structure of $tWb$ coupling, FCNC, fourth generation of the quarks and the existence of $W'$. 
\end{itemize}
\begin{table}[h]
\begin{center}
 \caption{The total cross sections for single top production processes.}
  \begin{tabular}{|l|ccc|} \hline
    process & t-channel & s-channel & tW-channel \\\hline
    cross section (pb) & 242.6(NLO) & 10.62(NLO) & 60(LO) \\ \hline
  \end{tabular}
    \label{tab:crosssection}
\end{center}
\end{table}
The total cross sections of single top production processes are listed in Table~\ref{tab:crosssection}. 
The t-channel process has the largest cross section, almost one third of $t\bar{t}$ cross section. The s-channel 
process has the smallest cross section, one order of magnitude less than t-channel
~\cite{singletop1,singletop2,singletop3,singletop4}.
The recent selection startegies for separation of signal from backgrounds for the three production processes
in ATLAS and CMS are described in the following sections.
\section{t-channel}
\begin{figure}[h]
\begin{center}
\epsfig{file=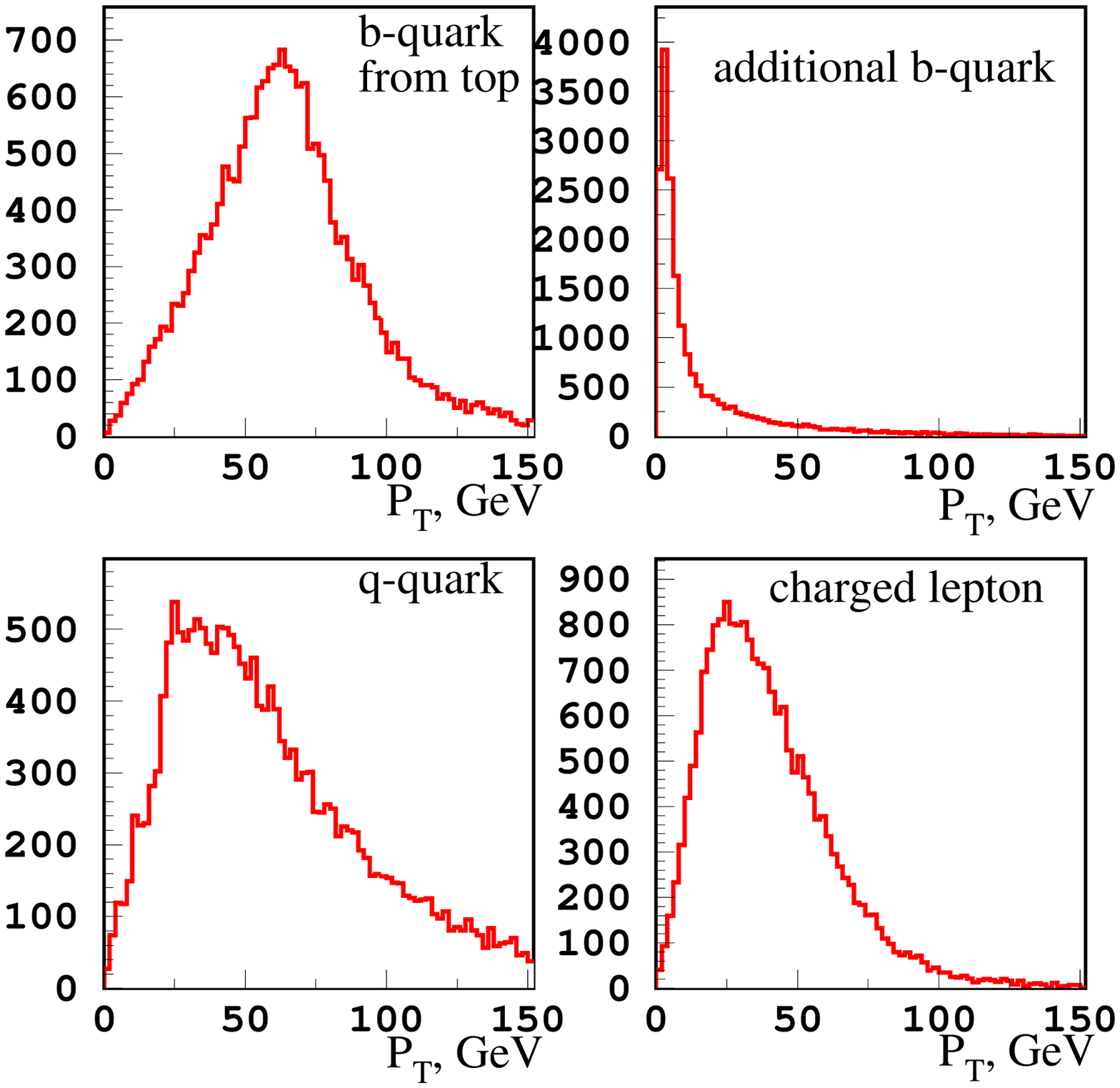,height=5cm,width=5cm}
\epsfig{file=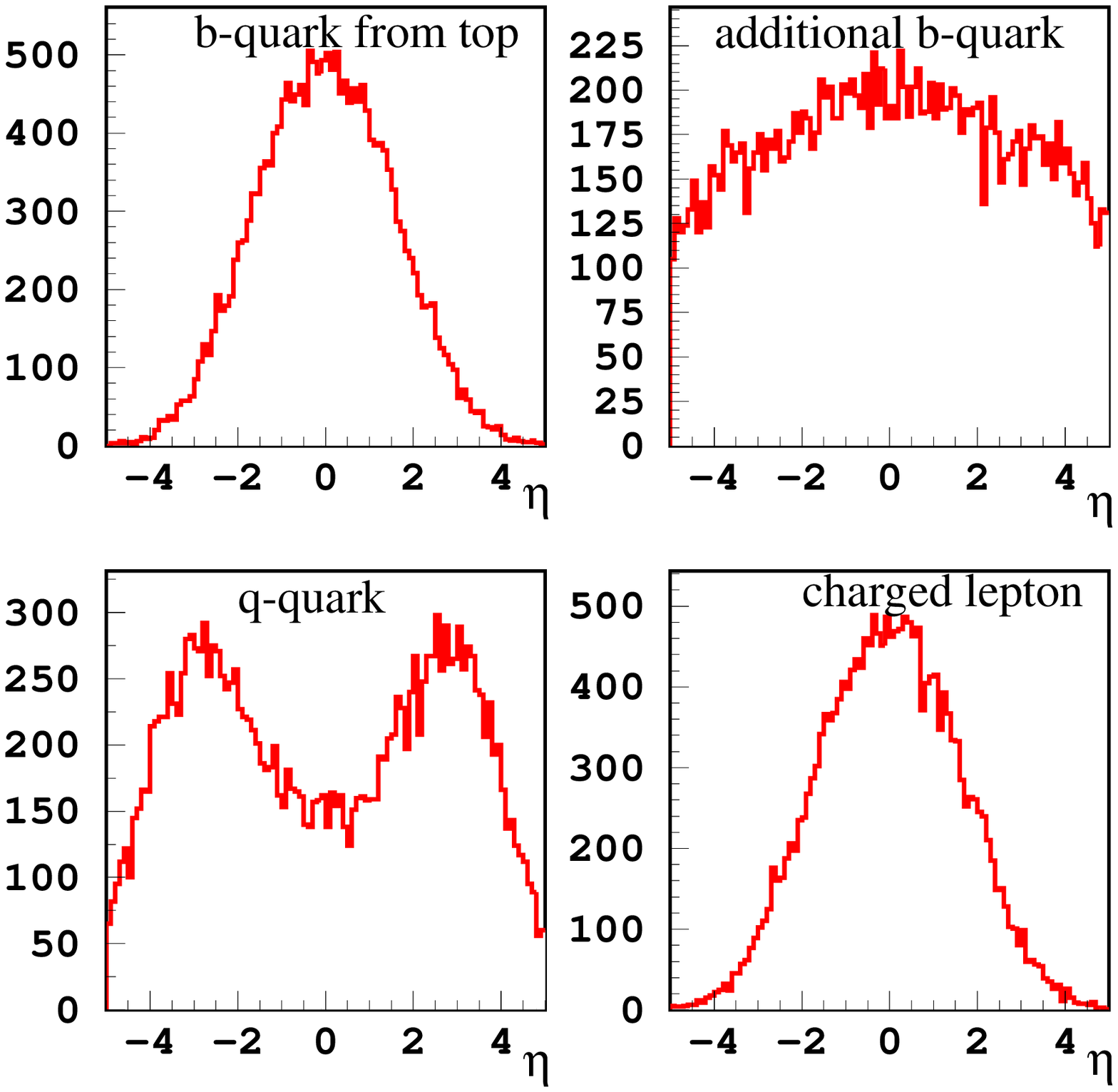,height=5cm,width=5cm}
 \caption{The transverse momenta and pseudo rapidity distributions of the final partons in t-channel.}
    \label{fig:TCH_part_Pt}
\end{center}
\end{figure}

In the t-channel process when the $W$ boson decays leptonically the final objects at parton
level are one b-quark from top, one charged lepton,  neutrino,
a light quark and an additional b-quark. The kinematical distribution of the final partons are shown 
in Fig.~\ref{fig:TCH_part_Pt}.
It is obvious from  Fig.~\ref{fig:TCH_part_Pt} that all final objects have relatively large transverse
momenta except for the additional b-quark which is very soft.
The identification (even the observation) of the additional b-quark is difficult due to its softness. 
The light quark tends to fly in the forward/backward direction which will be useful for separation of
signal from backgrounds.

Since detector can measure the transverse missing energy only,
one needs to calculate the longitudinal component of the neutrino. The longitudinal component
of the neutrino could be obtained by using the mass of the W boson and solving a quadratic equation:
\begin{eqnarray}
M_{W}^{2} = (P_{l} + P_{\nu})^{2}
\end{eqnarray}
This equation has two solutions:
\begin{eqnarray}
&P_{z,\nu}^{1,2} = \frac{AP_{z,l}\pm \sqrt \Delta}{P_{T,l}^{2}}\\ \nonumber
&A \equiv \frac{M_{W}^{2}}{2} + \vec{P}_{T,l}.\vec{E}_{T,miss}~,~\Delta \equiv E_{l}^{2}(A^{2} - E_{T,miss}^{2}P_{T,l}^{2})
\end{eqnarray}
Two difficulties are present here. First, in some events $\Delta < 0$, as a result no real solution exists 
for the lonitudinal component of neutrino. Second, there is an ambiguity in choosing the $P_{z,\nu}$ 
between the two roots.

Regarding the above properties of t-channel the selection strategies of 
ATLAS and CMS are presented below. Common selections are:
\begin{itemize}
\item One isolated high $p_{T}$ lepton in the central region.
\item Cut on missing transverse energy (useful for suppression backgrounds with $W$ boson in the final state).
\item Exactly two high $p_{T}$ jets, one of them should be b-tagged in centeral region and another one should be
in the forward region ($|\eta| > 2.5$).
\end{itemize}

In the ATLAS analysis, the events with no real solution for $P_{z,\nu}$ are discared and between two real solutions of $P_{z,\nu}$
, the solution which gives closest invariant mass of $l\nu b$ to the nominal top mass is chosen.
A window cut is applied on $H_{T}$ where $H_{T} = \Sigma_{jet} E_{T}^{jet}+E_{T,l}+E_{T,miss}$ and on the top mass.  
After following the above strategy and taking into account $t\bar{t}$ and $W+$jets backgrounds the
number of survived signal events are 7000 after 30 fb$^{-1}$ of integrated luminosity and $\frac{S}{B}$ 
and statistical sensitivity to $V_{tb}$ are:
\begin{eqnarray}
\frac{S}{B} \sim 3 ~~,~~ \frac{\sqrt{S+B}}{B} = 0.014 
\end{eqnarray}

\begin{figure}[h]
\begin{center}
\epsfig{file=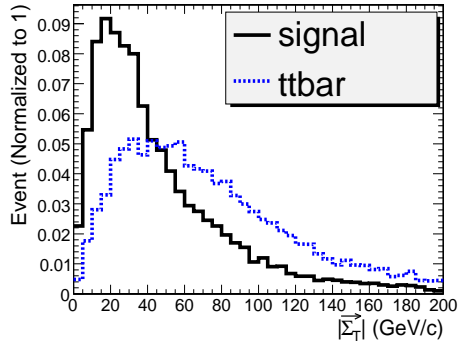,height=5cm,width=6cm}
 \caption{The distribution of $|\vec{\Sigma}_{T}|$ in t-channel (CMS).}
    \label{fig:TCHSigma.eps}
\end{center}
\end{figure}

The main dominant systematics are due to background estimation, ISR/FSR, JES, b-tagging and luminosity.
ISR and FSR have a significant influence on the jet multiplicity, the uncertainty on JES affects the selection efficiencies
directly(via the thresholds of the $p_{T}$ of the jets), the uncertainty on b-tagging has an important effect
on the background rejection(especially on rejection of $W+$njets).

In CMS analysis, the events with no real solution for $P_{z,\nu}$ are kept (in this case $M_{W}$ is increased up to the value in
which $\Delta = 0$ ) and between two real solutions of $P_{z,\nu}$,
the solution with minimal absolute value is chosen.
A cut is applied on $|\vec{\Sigma}_{T}|$ where $\vec{\Sigma}_{T} = \vec{P}_{T,W} + \vec{P}_{T,bjet} + \vec{P}_{T,forw.jet}$
(Fig.~\ref{fig:TCHSigma.eps}).
A window cut is also applied on transverse mass of the W boson where $M_{T}^{W} = \sqrt{ 2(P_{T,l} E_{T,miss} -
 \vec{P}_{T,l} \cdot\vec{E}_{T,miss})}$. Finally, a window cut is applied on the top mass.
At 10 fb$^{-1}$ and after applying the above strategy, the efficiency is around $1.4\%$ and number of survived 
events is around 2350 events.
\section{s-channel}
The final objects of the s-channel mode when $W$-boson decays leptonically are two high $p_{T}$
b-jets, one charged lepton and neutrino. In principle, searching for s-channel is more difficult
than t-channel because of much smaller cross section, the presence of two high $p_{T}$ b-jets, lack 
of any extra forward jet. However, the PDF's of the initial state of s-channel($u(\bar{u})$ and $d(\bar{d})$)
are very well known while the PDF's of the initial state of t-channel($b(\bar{b})$ and gluon) are less known,
so the uncertainty due to PDF in s-channel is smaller than t-channel.
In the following the selection strategies for CMS and ATLAS are presented.
Common selections are:
\begin{itemize}
\item One isolated energetic charged lepton in the centeral region.
\item A cut is applied on transverse missing energy.
\item Exactly two high $p_{T}$ b-tagged jet.
\end{itemize}
In the CMS analysis, the events with no real solution for $P_{z,\nu}$ are kept (in this case
the real part is kept) and between two real solutions of $P_{z,\nu}$,
the solution with minimal abstract value is chosen.
For top reconstruction, the b-jet which has the opposite charge sign ($Q_{jet}$) to lepton 
is used. Jet charge is defined as the sum of the charges of the tracks inside the jet cone. A cut 
is applied on $|\vec{\Sigma}_{T}|$ where $\vec{\Sigma}_{T} = \vec{P}_{T,W} + \vec{P}_{T,bjet1}+ \vec{P}_{T,bjet2}$.
A window cut is also applied on the top mass~\cite{Andrea}.  

\begin{figure}[h]
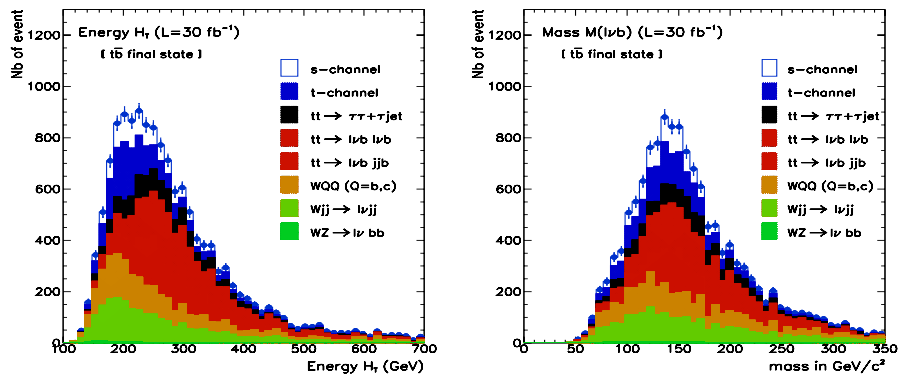

\begin{center}
\epsfig{file=figs/SCHHtATLAS.eps,height=5cm,width=6cm}
\epsfig{file=figs/SCHTopMassATLAS.eps,height=5cm,width=6cm}
 \caption{The distributions of $H_{T}$ and the top mass in the s-channel (ATLAS).}
    \label{fig:SCHTopMassATLAS.eps}
\end{center}
\end{figure}

In the ATLAS analysis, the selection is splitted in two parts: $t\bar{b}$ and $\bar{t}b$. The events containing
a secondary lepton with opposite sign are rejected. Among the four combinations 
of two b-tagged jets and two solutions for $P_{z,\nu}$,  
the combination which gives the closest invariant mass of $l\nu b$ to the nominal top mass is chosen.
A window cut is applied on $H_{T}$ where $H_{T} = \Sigma_{jet} E_{T}^{jet}+E_{T,l}+E_{T,miss}$
and also on the top mass (Fig.~\ref{fig:SCHTopMassATLAS.eps}).
The mentioned strategy for ATLAS leads to the following results at 30 fb$^{-1}$:
\begin{eqnarray}
\frac{S}{B} \sim 0.15 ~~,~~ \frac{\sqrt{S+B}}{B} = 0.06 
\end{eqnarray}
After application of all cuts the efficiency is $1\%$ and number of survived events is 2050.
The dominant background comes from the top pair production 
in the dilepton and lepton+jets channels, followed by $WQ\bar{Q} (Q = c,b)$ contamination.
The remaining $W+$jets contamination is due to the high cross section for such events.
The dominant systematics are systematics due to background estimation, ISR/FSR, JES, b-tagging and luminosity.
\section{tW-channel}
In ATLAS the study for tW-channel has been performed for the case of leptonic decay of the $W$-boson from top decay and 
hadronic decay of another $W$-boson. One high $p_{T}$ lepton and exactly three jets are required, one of jets should be
b-jet in the central region. $W$-boson is reconstructed by the two non b-tagged jets.
Window cuts are applied on $M_{W}$, the invariant mass of the $l\nu b$ system and on $H_{T}$.
After applying the cuts the signal to background ratio is $\frac{1}{7}$ and $\frac{\sqrt{S+B}}{B} = 0.04$ (at 30 fb$^{-1}$).
The dominant backgrounds are $t\bar{t}$ and $t$-channel. For an integrated luminosity of 30 fb$^{-1}$
ATLAS expects 4700 events in $tW$-channel. A large systematic uncertainty is expected to come from
the JES and from ISR/FSR.
\section{Conclusions}
Single top quark is expected to be observed in Tevatron soon. Since the cross sections of single top
quark production processes at LHC is much higher than Tevatron, LHC provides the possibility for precise
measurements of all three processes.
The electroweak model in the top quark sector could be tested and the $|V_{tb}|$ CKM matrix element is determined with high
precision. Single top production processes could be used as a powerful mean for probing new physics.
\section*{Acknowledgments}
I would like to thank to Arnaud Lucotte who provided the most recent updated results from ATLAS analyses.
Many thanks to the CMS single top group and Joachim Mnich.

\end{document}